\documentclass[twocolumn]{biophys}

\usepackage{helvet,times}
\usepackage{bm,textcomp}

\jno{kxl014}
\gridframe{N}
\cropmark{Y}
\doi{doi: }

\usepackage{amsmath,amssymb,amsfonts,amsthm}
\usepackage[round,numbers,sort&compress]{natbib}
\usepackage{siunitx}
\usepackage{tikz}

\DeclareSIUnit\molar{\mole\per\cubic\deci\metre}
\DeclareSIUnit\Molar{\textsc{m}}

\showboxbreadth\maxdimen\showboxdepth\maxdimen

\begin{document}

\setcounter{page}{1} 

\title{Steric Effects Induce Geometric Remodeling of Actin Bundles in Filopodia}

\author{Ulrich Dobramysl\thanks{Wellcome Trust/Cancer Research UK Gurdon Institute, University of Cambridge, Tennis Court Road, Cambridge CB2 1QN, United Kingdom; e-mail: {ulrich.dobramysl@gurdon.cam.ac.uk}.},
\quad
Garegin A. Papoian\thanks{Department of Chemistry \& Biochemistry, University of Maryland, Chemistry Bldg 2216, College Park, MD 20742-2021, United States; e-mail: {gpapoian@umd.edu}.}$\,\;$\thanks{G.A.P. and R.E. are joint corresponding authors.}
\quad and \quad
Radek Erban\thanks{Mathematical Institute, University of Oxford, Radcliffe Observatory Quarter, Woodstock Road, Oxford, OX2 6GG, United Kingdom; e-mail: {erban@maths.ox.ac.uk}.}$\,\;^\ddagger$}


\begin{abstract}%
{Filopodia are ubiquitous finger-like protrusions, spawned by many eukaryotic cells, to probe and interact with their environments. Polymerization dynamics of actin filaments, comprising the structural core of filopodia, largely determine their instantaneous lengths and overall lifetimes. The  polymerization reactions at the filopodial tip require transport of G-actin, which enter the filopodial tube from the filopodial base  and diffuse towards the filament barbed ends near the tip. Actin filaments are mechanically coupled into a tight bundle by cross-linker proteins. Interestingly, many of these proteins are relatively short, restricting the free diffusion of cytosolic G-actin throughout the bundle and, in particular, its penetration into the bundle core. To investigate the effect of steric restrictions on G-actin diffusion by the porous structure of filopodial actin filament bundle, we used a 	particle-based stochastic simulation approach. We discovered that excluded volume interactions result in partial and then full collapse of central filaments in the bundle, leading to a hollowed-out structure. The latter may further collapse radially due to the activity of cross-linking proteins, hence producing conical shaped filament bundles. Interestingly, electron microscopy experiments on mature filopodia indeed frequently reveal actin bundles that are narrow at the tip and wider at the base. Overall, our work demonstrates that excluded volume effects in the context of reaction-diffusion processes in porous networks may lead to unexpected geometric growth patterns and complicated, history-dependent dynamics of intermediate meta-stable configurations.}%
{Received}%
{Correspondence:}
\end{abstract}

\maketitle

\markboth{Dobramysl et al.}{Steric Effects Induce Remodeling of Filopodia}


\section{INTRODUCTION}
Many eukaryotic cells project dynamic finger-like protrusions, called filopodia, that are comprised of a bundle of actin filaments enveloped by the cellular membrane~\cite{ref:mattila08,ref:small10}. Filopodia play diverse roles across many cell types. In particular, signaling via receptors on filopodial tips allows cells to sense their environment and guide  chemotaxis~\cite{ref:han02}. Neurons use filopodia in axonal growth cones, to determine the direction of elongation and branching~\cite{ref:dent03}, as well as in dendritic spine formation~\cite{ref:maletic99}. During wound healing, ``knitting'' of filopodia protruding from epithelial cells plays an important role~\cite{ref:wood02}. Filopodia are also implicated in cancer progression and metastasis because of their involvement with cell motility~\cite{ref:arjonen11}. They also arise in some viral infections, creating physical connections among the hosts' cells~\cite{ref:sherer07}.

G-actin, an abundant and highly conserved protein, self-assembles into double helical filaments, called F-actin. The latter is the fundamental building block of eukaryotic cellular cytoskeletons. F-actin structure is polarized, with polymerization at the barbed end being more efficient by an order of magnitude compared to the pointed end, while having similar depolymerisation rates at both ends. This asymmetry, based on the hydrolysis of ATP into ADP by actin molecules, leads to ``treadmilling'', whereby filaments can convert chemical energy stored in ATP into mechanical work of pushing against the external resistance. Hence, actin filaments are dynamic, dissipative structures, that allow for fast morphological transitions in the cellular cytoskeleton in response to external and internal biochemical and mechanical cues, mediated by a vast array of signaling and regulatory proteins. 
\begin{figure}[tb]
  \centering
  \centerline{\includegraphics[width=0.99\columnwidth]{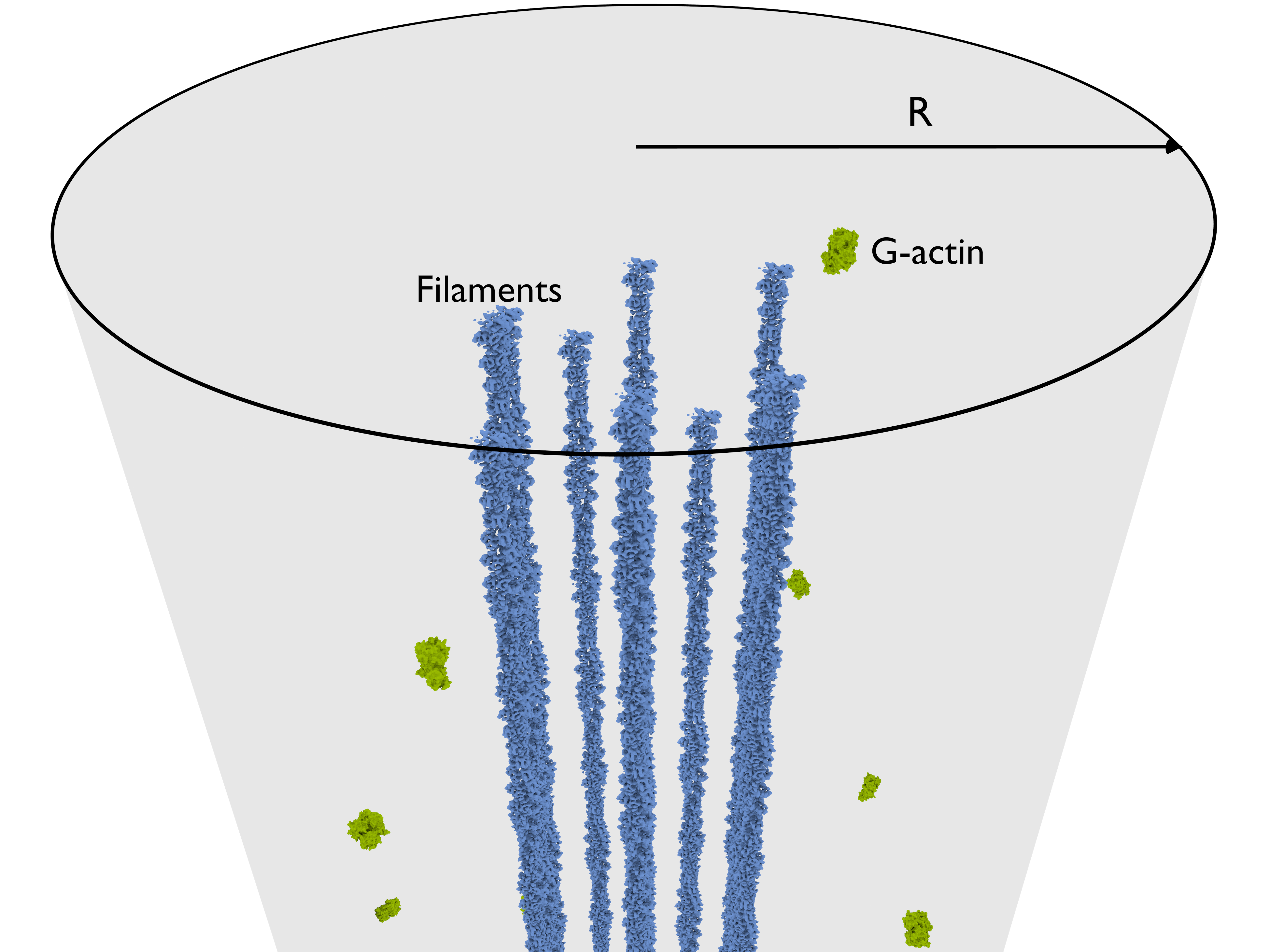}}
  \caption{\label{fig:filopodia_sketch} 
  {\it Rendering of an actin filament bundle
    inside a filopodium. The filaments are shown in blue, with diffusing G-actin
    monomers displayed in green. The gray region indicates the volume enclosed
    by the membrane in our model. The filament height and G-actin position data
    stem from an example simulation run. Filament and G-actin structure data
    were taken from the PDBe database}~\cite{Behrmann2012,Galkin2008}.}
\end{figure}

In filopodia, which are roughly cylindrical tubes with radius
$R \sim 50-250\operatorname{nm}$~\cite{Atilgan2006,ref:sheetz92}, between 10 and
30 actin filaments are organized into parallel, tightly cross-linked structures
(see Fig.\ \ref{fig:filopodia_sketch}). The filopodial lengths, $L$, vary
considerably between cell types, ranging from a few up to several tens of
microns~\cite{ref:mogilner05,ref:bray01}. Many proteins are involved in the
complex machinery responsible for the formation and subsequent biological
function of filopodial protrusions, including formins~\cite{Hotulainen2006},
ENA/VASP~\cite{Bear2002} and capping proteins~\cite{Schafer1996}, among others,
however, how these proteins regulate the growth-retraction dynamics of filopodia
is still not fully
understood~\cite{ref:mattila08,ref:evangelista03,Faix2009,Leijnse2015}.

In a series of works, Papoian and coworkers have proposed a theory for length
regulation in filopodia~\cite{ref:lan08,ref:zhuravlev09}. For a stationary
filopodium, the key idea is to match three actin fluxes: the consumption of
G-actin at the filopodial tip, $J_p$, must equal the transport or diffusional
flux of G-actin to the tip, $J_d$, and the flux of actin subunits leaving the
filopodial tube, $J_r$, due to retrograde flow. The latter process is
mechanically mediated by the polymerization of actin against the resistance of
the filopodial membrane and, to a larger extent, by contractile dynamics of
actin network inside the cell body, to which the roots of filopodial actin
bundles are
anchored~\cite{Leijnse2015,Mejillano2004,Heid2005,Mallavarapu1999a,Aratyn2007,Bornschlogl2013}. In
the mean-field limit, the equations, $J_p = J_d = J_r$, can be solved to yield a
closed form expression for the stationary length of
filopodia~\cite{ref:lan08}. Prior models also investigated filopodial growth
dynamics, using somewhat different assumptions and computational
approaches~\cite{ref:mogilner05,Atilgan2006,Erban2013}. Detailed stochastic models of
filopodial growth that included important regulatory proteins such as capping
proteins, resulted in macroscopic length fluctuations of filopodial protrusions
reminiscent of ubiq\-ui\-tous\-ly observed filopodial growth-retraction
cycles~\cite{ref:zhuravlev09}. These near critical length fluctuations arise
from the amplification of molecular noise due to binding and unbinding of
capping proteins to filament ends~\cite{ref:zhuravlev09}. In addition, Zhuravlev
and Papoian found that capping proteins induce bundle thinning, leading to the
eventual full collapse of filopodia, suggesting a microscopic mechanism for
filopodial aging and limits on their lifetime~\cite{ref:zhuravlev09}.

Excluded volume effects have been studied in the context of macromolecular
crowding in cells. It has been shown that crowding introduces a size-dependent
viscosity inside the cytoplasm~\cite{Kalwarczyk2011}, rendering the motion of
proteins highly non-trivial. The complex structure of chromatin inside cell
nuclei leads to a significant reduction of the effective diffusion coefficient
of fluorescent probes~\cite{Wachsmuth2000} thereby possibly influencing and
regulating the transcription of DNA into mRNA and the subsequent translation
into proteins. Simulation studies and analytical work considering volume
exclusion done on chromatin structure extracted from real nuclei via x-ray
tomography showed that there exists an optimal amount of volume exclusion that
minimizes the mean time for DNA-binding proteins to find their binding
sites~\cite{Isaacson2013}. Theoretical works have shown that Brownian motion
with hard-sphere interactions can be captured via an effective, collective
diffusion constant~\cite{Bruna2012}. Considerably less work has been done to
study the feedback between growing cytoskeletal structures and the resulting
restriction of the free diffusion of their building blocks.

Here, we show that excluded volume effects also result in filament collapse and
bundle thinning. However, in contrast to the effect of capping protein, steric
constraints on G-actin diffusion in the bundle interior result in geometrically
preferential growth-retraction patterns. Instead of any filament in a bundle
being susceptible to collapse due to capping, core filaments in the bundle are
starved of G-actin and eventually collapse. G-actin monomers are proteins with
an approximately ellipsoidal shape, with the semi-principal axes of
$\num{6.7}\times\num{4}\times\SI{3.7}{\nano\meter}$~\cite{Suck1981}. In
addition, both cytosolic and filamentous actin molecules are negatively charged,
resulting in their mutual electrostatic repulsion. Because a typical
electrostatic screening length is on the order of 1 nm under physiological
conditions~\cite{Spitzer2009}, the electrostatic repulsion is expected to add
approximately 2 nm to the steric exclusion zone~\cite{Evans1999} when
considering penetration of G-actin molecules in-between two F-actin
filaments. In our calculations, we consider G-actin as spherical particles with
a radius of $\SI{3.5}{\nano\meter}$, averaging over its ellipsoidal semi-axes
and also taking into account the screened electrostatic repulsion between
G-actin and F-actin. The diameter of actin filaments is approximately
$\SI{7}{\nano\meter}$~\cite{Tucker1971,Heath1978,Fujii2010}. The average
inter-filament spacing inside a bundle of actin filaments is
$\SI{12}{\nano\meter}$~\cite{Jasnin2013,Urban2010,Yang2013} with a range of
about $\num{10}-\SI{13}{\nano\meter}$~\cite{Yang2013}. In terms of lateral
geometrical placements, actin filaments in a filopodial bundle were found to be
arranged on an ordered hexagonal lattice~\cite{Jasnin2013}. Therefore, the
geometry of the porous networks found in these tightly cross-linked bundles is
likely to impede monomeric G-actin passing through the space between two
neighboring filaments. The resulting hindered diffusion of G-actin restricts
availability of G-actin in the bundle interior.

Our simulations show that the above outlined steric exclusion has profound
effects on the shape of the filopodial tip. The initial transient growth phase,
during which some interior filaments collapse, results in a stationary filament
length configuration that is only meta-stable. Over time, interior filament
height fluctuations drive them below a mean-field cut-off height (discussed
below), when they subsequently collapse to a new meta-stable configuration at an
intermediate height and eventually collapse completely. The partial or full
collapse of an interior filament creates a new diffusion channel for G-actin to
explore. These channels determine the subsequent critical lengths for the
fluctuations of the remaining interior filaments. Hence, the further evolution
of the inner filament heights is highly dependent on the time ordering of
previous filament collapses, leading to history-dependent meta-stable filopodial
states. Over time, these processes completely hollow out the interior of actin
bundle at the filopodial top, while leaving the peripheral filaments largely
intact. In terms of long-term evolution, such a hollow structure may not be
stable with respect to the activity of cross-linking and motor proteins, leading
to a global geometric reshaping of the filopodial actin bundle, which, in turn,
would significantly diminish its mechanical stability, as discussed below.

In the next section, we discuss our methodology and tools. In
Section~\ref{sec:results}, we present our simulation results and discuss the
most salient meta-stable states arising from the volume exclusion effects. In
Section~\ref{sec:mean-field-model}, we introduce a mean-field model that allows
us to gain further insight into the stability conditions of the observed states,
finding that the mean-field predictions for the heights of partially collapsed
filaments closely agree with the Brownian dynamics simulation results. Finally,
we analyze the mechanical implications of the morphological transitions induced
by the above-mentioned excluded volume effects.

\begin{table}[t]
  \caption{\label{tab:parameters} 
 Parameter values used throughout this study. The numbers in the 
third column indicate references for the cited parameter values.}
 \vskip 1mm 
 \begin{tabular}{lll}
    \hline
    Description & Value & Ref. \\
    \hline
    \multicolumn{3}{l}{\rule{0pt}{10pt}G-actin parameters} \\
    ~Diffusion constant & $D\!=\!5\!\times\!10^6\,\si{\nano\meter\tothe{2}\per\second}$\hspace{-8pt} & \cite{ref:mogilner05} \\
    ~Bulk G-actin & $c_0\!=\!\SI{10}{\micro\Molar}$ & \cite{Pollard2000} \\
    \multicolumn{3}{l}{\rule{0pt}{10pt}Geometry} \\
    ~Filopodial radius & $R\!=\!\SI{75}{\nano\meter}$ & \cite{Atilgan2006} \\
    ~Filament spacing & $d\!=\!\num{11},\num{12},\SI{13}{\nano\meter}$ & \cite{Jasnin2013,Urban2010,Yang2013} \\
    ~Filament radius & $r_F\!=\!\SI{3.5}{\nano\meter}$ & \cite{Tucker1971,Heath1978,Fujii2010} \\
    ~Filament number & $N\!=\!19$ & \cite{ref:sheetz92} \\
    \multicolumn{3}{l}{\rule{0pt}{10pt}Filament dynamics}\\
    ~Binding radius & $\varrho\!=\!2r_F$ & \\
    ~Actin on rate & $k^+\!\!=\!\SI{11.6}{\micro\Molar\tothe{-1}\second\tothe{-1}}$\hspace{-8pt} & \cite{Pollard1986,Fujiwara2007} \\
    ~Actin off rate & $k^-\!\!=\!\SI{1.4}{\second\tothe{-1}}$ & \cite{Pollard1986,Fujiwara2007} \\
    ~Retrograde flow & $v\!=\!\SI{70}{\nano\meter\per\second}$ & \cite{ref:lan08} \\
    ~Polymer. length & $\delta\!=\!\SI{2.7}{\nano\meter}$ & \cite{ref:lan08} \\
    \multicolumn{3}{l}{\rule{0pt}{10pt}Membrane interaction} \\
    ~Membrane force & $f\!=\!\SI{10}{\pico\newton}$ & \cite{ref:lan08,Peskin1993} \\
    ~Fluctuation size & $\sigma\!=\!\SI{20}{\nano\meter}$ & \cite{ref:lan08,Peskin1993} \\
    ~Temperature & $k_BT\!=\!\SI{4.1}{\pico\newton\nano\meter}$ & \cite{ref:mogilner05} \\
    \hline
  \end{tabular}
\end{table}

\section{MATERIALS AND METHODS}
In order to investigate how steric interactions among filaments and G-actin
monomers affect actin polymerization dynamics, we have developed a simulation
model that incorporates excluded volume effects. In our model, we allow the
elongation of actin filaments via binding of G-actin molecules. G-actin
particles are allowed to move according to Brownian Dynamics (BD, see below) and
can bind to microfilaments when they enter the vicinity of a binding site on top
of a filament, thereby elongating the filament by a length $\delta$. Note that
the real mechanism of actin polymerization at the filopodial tip may be more
complicated and involves additional protein complexes such as ENA/VASP and
formins. Therefore our model should be treated as effectively averaging over
these details with an effective (or renormalized) polymerization rate. The
simulation model proceeds via discrete time steps
$\Delta t=\SI{100}{\nano\second}$, in which molecules first move via the BD
step, subsequently molecules near binding sites are allowed to bind, and finally
G-actin can depolymerize from the tips of filaments, freeing a molecule and
reducing the filament length by $\delta$. The parameter values used in this
study are summarized in Table~\ref{tab:parameters}. We ran 200 distinct
realizations of this model for 2,000 seconds each to extract filament height
trajectories over time for three different values of the interfilament spacings
(taking approximately 50,000 CPU hours).

\subsection{Particle-based stochastic model for diffusion}
Our spatially-extended stochastic model for filopodial growth is confined to a
spatial domain with cylindrical shape of variable length
$L(t)=L_F(t)+\SI{25}{\nano\meter}$ (depending on the length of the filopodium at
a time $t$, which is given by the highest filament at the $L_F(t)$) and radius
$R=\SI{75}{\nano\meter}$. G-actin monomers can enter and exit the domain via the
boundary at the bottom, which is held at a constant concentration. These
particles undergo BD with a fixed time step $\Delta t$, i.e. their position is
updated according to
$ \mathbf{X}_{j}(t+\Delta t) = \mathbf{X}_{j}(t) + \sqrt{2D\Delta t}\,
\boldsymbol{\xi}_{j}$
with $j=1,2,\dots,N_A(t)$, where $N_A(t)$ is the number of G-actin molecules at
a given time $t$, and $\boldsymbol{\xi}_{j}$ is a vector of independent normally
distributed random numbers with zero mean and unit variance. Volume exclusion of filaments and boundaries are handled via
reflection along the surface normal.

\subsection{Volume exclusion}
Filaments are modeled as rigid cylinders with a finite radius $r_F$ placed in a
hexagonal arrangement (see Section~\ref{sec:latpos}). To account for the spatial
extent of G-actin molecules (which is assumed to be spherical, with the same
radius $r_F$), the effective radius of filaments is given as $2r_F$ (and
diffusing molecules are implemented as points). To facilitate this study,
G-actin molecules do not mutually interact, hence we do not take queuing and
crowding effects into account as this would become computationally
prohibitive. The concentration of G-actin is relatively small, hence we do not
expect large effects from neglected interactions of G-actin molecules.

\subsection{Molecule binding}
In order to model the binding of molecules to filament tips, we implement a
reversible binding scheme~\cite{Lipkova2011}, with a binding radius of
$\varrho=2r_F=\SI{7}{\nano\meter}$. In this scheme, molecules become binding
candidates as soon as they enter a sphere of radius $\varrho$ around the tip of
a filament. They are then allowed to bind to the site with a probability
$P_\lambda$ per time step that they spend in the binding
region~\cite{Lipkova2011,Erban:2009:SRD}. In the vicinity of the top of the
filopodium, the binding probability is modified to take the force enacted on the
bundle via the membrane into account (see below). We calculate the remaining
parameter $P_\lambda$ (binding probability) before the start of simulations
using the approach described in~\cite[Section 5]{Lipkova2011}. To implement
depolymerization, a molecule is allowed to dissociate from a filament tip with a
probability $1-\exp(- k^- \, \Delta t)$ and placed at the former position of the
tip. Here, we neglect possible changes in the depolymerization rate due to the
hydrolysis of actin molecules as the resulting large fluctuations in filament
length occur only at actin concentrations much smaller than exist in our
simulations~\cite{Vavylonis2005}.

\subsection{Membrane force}
\label{sec:membraneforce}
Experiments have shown that the actin polymerization rate highly depends on the
force enacted on filaments by the membrane~\cite{Carlsson2001}. This effect can
be thought of as temporary envelopment of a filament tip by the membrane,
sterically disallowing the binding of G-actin to the tip, thereby synchronizing
the movement of the filament tips close to the membrane~\cite{ref:lan08}. Thus,
the bare bulk polymerization rate $k^+$ needs to be multiplied by the
probability for the creation of enough space between a filament tip and the
membrane to accommodate the molecule. This Brownian ratchet model allows us to
take into account membrane effects without explicitly simulating the movement of
the membrane by modifying the binding probability of G-actin molecules inside
the binding radius of the $i$-th filament~\cite{ref:zhuravlev09,Peskin1993}
\begin{equation}
  \label{eq:membrane-rate}
  P_{\lambda}^{(i)}=P_\lambda\exp\left(-\frac{f_i\delta}{k_BT}\right)\;,\qquad i=1,2,...,N\;,
\end{equation}
where $f_i$ is the force on filament $i$.
On the micrometer scale, the cell membrane is known to equilibrate on the order
of microseconds~\cite{Lin2004}, hence much faster than the average time between
polymerization reactions. Therefore, $f_i$ is
proportional to the probability of the filament tip being covered by the
membrane. Assuming Gaussian-distributed membrane height fluctuations, we can
follow Lan and Papoian~\cite{ref:lan08} to calculate the probability of the
membrane being below the filament tip
\begin{equation*}
  p_i=
  \frac{1}{2}\mathrm{erfc}\left(\frac{\max\{h_j\}-h_i}{\sigma}\right)\;,
\end{equation*}
where $\mathrm{erfc}(x)=1-\mathrm{erf}(x)$ is the complementary error function
and $h_j$ is the height of the $j$-th filament.  The total membrane force $f$ is
distributed across all the filaments according to this probability, hence the
force on filament $i$ is
$$
f_i= \frac{p_i f}{\sum_{j=1}^{N}p_j},
$$ 
which we can substitute into Eq.~(\ref{eq:membrane-rate}) to get the effective
polymerization probability.

Wang and Carlsson~\cite{Wang2014} model the motion of a Brownian obstacle (the
membrane in our case) pushed by polymerizing filaments. They disallow
polymerization when the distance to the membrane is less than the monomer size
and let polymerization proceed with the bare rate $k^+$ otherwise. Das {\it et
  al.}~\cite{Das2014} consider a discrete lattice model for filament
polymerization against an external load that reduces the polymerization rate of
the leading filament (in contact with the obstacle) only. Our model differs from
both of these in that we assume a non-rigid, highly dynamic membrane as the
obstacle, which can affect the polymerization of filaments even a distance below
the mean membrane position~\cite{ref:lan08}.

\subsection{Filament polymerization and collapse}
The dynamics of actin filaments are governed by the inherent asymmetry between
the two ends of a filament. The barbed end (the filament tip in our simulations)
exhibits a much higher polymerization rate, which leads to a tread-milling
process. Actin microfilaments grow via the binding of G-actin monomers, which
enhances the filament length by $\delta=\SI{2.7}{\nano\meter}$. The binding
algorithm implemented in our simulations is discussed above. The bulk
polymerization rate is given by
$k_A^+=\SI{11.6}{\micro\Molar\tothe{-1}\second\tothe{-1}}$ from which we
calculate the binding probability for a given binding radius. The binding radius
cannot be larger than half of the distance between filaments such that the
binding region does not overlap with excluded volume regions, thereby
artificially lowering the effective binding rate. During a depolymerization
reaction, the filament's height is reduced by a length $\delta$ and a new
G-actin molecule is placed at the former position of the filament tip. Enhanced
depolymerization processes at the spiked ends of actin filaments in the
lammellipodium lead to a pulling back motion of filaments called retrograde
flow. We incorporate this in our simulation model via a constant and negative
effective velocity of filament tips $v$. During each time step, each filament's
height shrinks by $v\Delta t$, in addition to any depolymerization reaction
occurring during this time step.

\subsection{Lateral position of filaments and initial state}
\label{sec:latpos}
Filaments are placed in a hexagonal arrangement with an initial height of
$\SI{100}{\nano\meter}$. The in-plane coordinates $\vec{x}_i$ of the $N=19$
filaments are given by
\begin{equation*}
  \begin{split}
    &(0,0), \quad (0,d), \quad (\sqrt{3}d/2,d/2), \quad (\sqrt{3}d/2,-d/2), \\
    &(0,-d), \quad (-\sqrt{3}d/2,-d/2), \quad (-\sqrt{3}d/2,d/2), \\
    &(\sqrt{3}d/2,3d/2), \quad (\sqrt{3}d,0), \quad (\sqrt{3}d/2,-3d/2), \\
    &(-\sqrt{3}d/2,-3d/2), \quad (-\sqrt{3}d,0), \quad (-\sqrt{3}d/2,3d/2), \\
    &(0,2d), \quad (\sqrt{3}d,d), \quad (\sqrt{3}d,-d), \quad (0,-2d), \\
    &(-\sqrt{3}d,-d), \quad (-\sqrt{3}d,d),
  \end{split}
\end{equation*}
with $d$ the next-neighbor distance. The hexagonal arrangement of microfilaments
in a bundle has been observed in experiments~\cite{Jasnin2013} and is close to an optimal
cross-sectional configuration for a bundle of actin filaments, maximizing an
individual filament's number of next neighbors.

\section{RESULTS}
\label{sec:results}
\begin{figure}[tb]
  \centering
  \includegraphics[width=0.99\columnwidth,page=1]{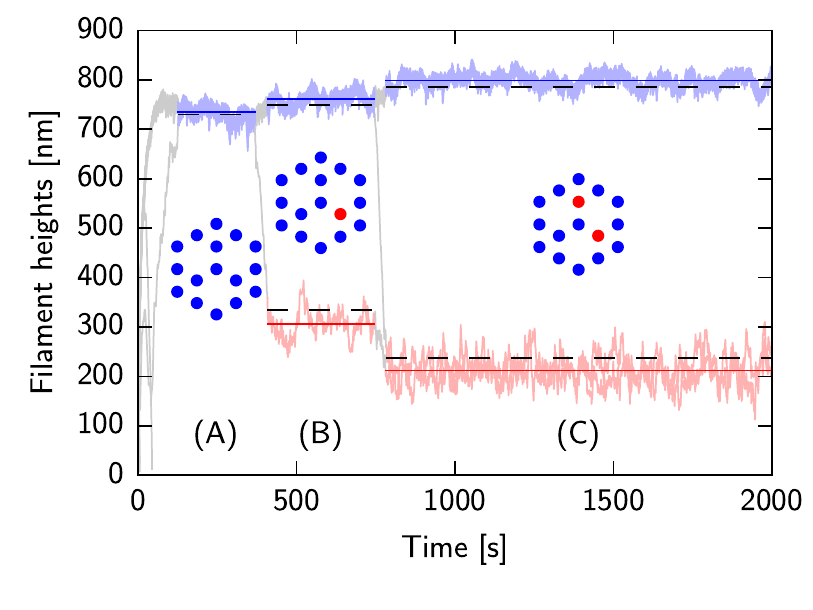}
  \caption{\label{fig:example-config} 
  {\it Filament heights over time from a
    representative simulation run with {\normalsize
      $d=\SI{13}{\nano\meter}$}. The upper filament trajectories at the bundle
    tip are highlighted in blue, the partially-collapsed filament trajectories
    are shown in red. All filaments are accounted for in these two categories
    during an identified meta-stable state. Three distinct meta-stable states
    are visible in order of appearance in the graph:} (A) 
    {\it Three completely
    collapsed filaments with the remainder at the tip of the bundle;} 
    (B) {\it one of
    the inner filaments collapsed partway;} 
    (C) {\it two inner filaments collapsed
    partway. The insets show the spatial configuration of the meta-stable
    states. The black dashed lines indicate the filament heights calculated from
    Eq.~$(\ref{eq:mf-stabilitycond})$. Gray lines indicate transient states
    between meta-stable configurations.}}
\end{figure}
\begin{figure}[tb]
  \centering
  \centerline{\includegraphics[width=0.99\columnwidth,page=2]{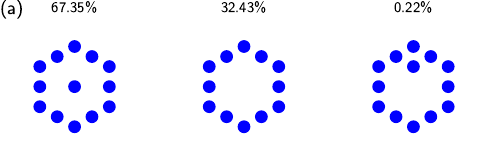}}
  \centerline{\includegraphics[width=0.99\columnwidth,page=3]{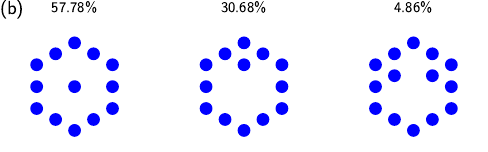}}
  \centerline{\includegraphics[width=0.99\columnwidth,page=4]{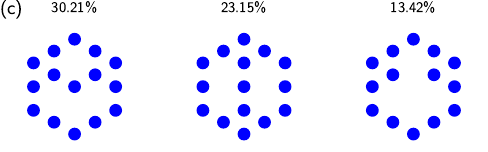}}
  \caption{\label{fig:metastable-states} 
  {\it The spatial distribution of full-length
    filaments in the actin filament bundle inside the filopodium, displayed for
    the $3$ most prevalent configurations out of $16$ found for}
    (a) {\normalsize
      $d=\SI{11}{\nano\meter}$}, (b) {\normalsize $d=\SI{12}{\nano\meter}$} 
      {\it and}
    (c) {\normalsize $d=\SI{13}{\nano\meter}$}.
    {\it A blue color indicates
    full-length filaments, while white space indicates partially or fully
    collapsed filaments. The numbers above the individual states indicate the
    percentage of total simulation time spent in this state.}}
\end{figure}

When filopodial growth dynamics are simulated using the approach described
above, we find that, typically, an actin filament bundle undergoes several
transitions between meta-stable configurations. Initially, a rapid growth phase
is observed, during which all filaments grow together, followed by intermittent
collapses of filaments {\it inside} the bundle. Hence, the bundle settles into
different meta-stable steady states, during which the filament heights fluctuate
around particular quasi-stationary values. These quasi-stationary heights are
determined by a balance between the polymerization of G-actin and the
combination of depolymerization of F-actin and retrograde flow (see
Section~\ref{sec:mean-field-model})~\cite{ref:lan08,Zhuravlev2011}. At
intermediate times, large fluctuations of single filaments push the bundle out
of the current meta-stable state and one or more filaments further
collapse. This collapse can be complete (a filament's height shrinks to zero and
it disappears) or partial. In the case of a partial collapse, collapsed
filaments fluctuate around a new, lower height, which is largely determined by a
changed supply of G-actin molecules due to the new bundle geometry. Hence, the
bundle height transitions due to individual filament collapses are exclusively
driven by the fluctuations of G-actin's availability mediated by local steric
constraints. This is in contrast to filament collapse due to the binding of
capping proteins that instead directly halt filaments' polymerization
propensities~\cite{ref:zhuravlev09}. Note that the polymerization reaction is
always diffusion limited, because the filaments will simply grow until a balance
is reached between the addition of new monomers and the reduction in length due
to depolymerization and retrograde flow~\cite{ref:lan08}. Hence, our results are
robust even for greater G-actin concentrations or large uncertainties in the
polymerization rate parameter. We checked this with sets of simulations with
$c_0=\SI{20}{\micro\Molar}$, $k^{+}=\SI{5}{\per\micro\Molar\per\second}$ and
$k^{+}=\SI{20}{\per\micro\Molar\per\second}$ (see below).

Fig.\ \ref{fig:example-config} shows filament height dynamics from a
representative simulation run. The height data clearly reveal three distinct
meta-stable states, which were automatically identified using the DBSCAN
clustering algorithm~\cite{Ester1996,scikit-learn}. Uncategorized data are shown
as light gray lines, while colored line plots show filament height data
belonging to an identified meta-stable configuration. The insets show a top view
of the spatial distribution of filaments inside the bundle. Blue circles
indicate the lateral positions of filaments in the cluster at the filopodial
top; red circles are the positions of filaments that are part of an
intermediate-height cluster; and white spaces indicate the former positions of
filaments that collapsed completely. Importantly, these empty spaces serve as
additional channels for G-actin to diffuse inside the bundle.

After the brief initial transient, Fig.\ \ref{fig:example-config} shows that
three filaments have completely collapsed and disappeared. The resulting bundle
is found, therefore, in a highly symmetric configuration with all remaining
filament tips at the top of the filopodium. Subsequently, at
$t\approx\SI{400}{\second}$, a large fluctuation drives a single filament tip
below the stable region and it collapses towards a level of about
$\SI{300}{\nano\meter}$. Note that at the same time, the filopodial tip
(i.e. the filament cluster at the tip of the filopodium) grows slightly. At the
second transition point, at $t\approx\SI{750}{\second}$, a second filament
collapses partway and joins the lower cluster. The two filament's height shrinks
slightly while the filopodial tip grows again. The occurrence of these
meta-stable states, together with the transitions between them strongly
indicates the existence of multi-stability in this system. When filament tips
inside the bundle leave the region of attraction of a stable cluster, they
collapse partway and reach a different meta-stable state.

Over time, this leads to a hollowing out of the filament bundle, which, in turn,
has interesting implications for the mechanical and geometrical properties of
the filopodium. Fig.\ \ref{fig:metastable-states} shows the three most prevalent
meta-stable configurations (considering only the filaments that are positioned
at the top of the bundle), ranked by the percentage of time spent in this state
throughout our study (we accounted for symmetries that make configurations
equivalent) for a filament spacing of $d=\num{11},\num{12}$ and
$\SI{13}{\nano\meter}$ (as this covers the majority of interfilament distances
observed in experiments~\cite{Yang2013}). Blue circles represent filaments that
reach the top of the filopodium, while white spaces indicate completely or
partially collapsed filaments (i.e. open diffusion channels inside the
bundle). All of these configurations show a thinning of the filament bundle
towards the filopodial top. As a general rule, when filaments in the bundle are
positioned with smaller interfilament spacings with respect to each other, the
inner filaments are more likely to collapse.

Because the parameter values listed in Table~\ref{tab:parameters} might have
large associated uncertainties or might apply only to a limited range of cell
types, we tested whether our observed effects are present also for different
parameter values. We systematically varied the following parameters:
\begin{itemize}
\item {\bfseries Bulk G-actin concentration:} The reported G-actin concentration
  measured in various cell types and organisms ranges between $\num{10}$ to
  $\SI{100}{\micro\Molar}$ and sometimes even
  higher~\cite{Pollard2000,ref:zhuravlev12}. However, for the purposes of our
  simulation study it is crucially important to distinguish not only between
  monomeric G-actin and polymerized F-actin, but also between G-actin that is
  available to polymerize and sequestered G-actin (i.e. bound to other proteins
  such as cofilin or thymosin-$\beta4$). The latter is unavailable to directly
  polymerize into filaments und therefore needs to be excluded. From
  quantitative simulation studies, Mogilner and
  Edelstein-Keshet~\cite{Mogilner2002} estimate the available unsequestered
  G-actin in the lamellipodium to be $\approx\SI{14}{\micro\Molar}$ when the
  total available actin concentration was $\SI{250}{\micro\Molar}$. From this,
  Zhuravlev {\it et al.} estimate a valid unsequestered G-actin concentration
  ranging between $\num{1}-\SI{50}{\micro\Molar}$~\cite[Supporting table
  1]{ref:zhuravlev12}.

  To check if our results are valid for higher concentrations within this range,
  we ran simulations for $c_0=\SI{20}{\micro\Molar}$, leading to a roughly
  doubled length of the filopodium in which the collapse of inner filaments
  together with meta-stable states still occur. We do not expect
    this to change for even higher G-actin concentrations, because inner
    filaments are always at a disadvantage due to the lower supply of G-actin
    monomers as long as bundles are able to grow when polymerization of
    filaments outperforms depolymerization and retrograde flow:
    \[c_0 k^+ > k^-+v_{ret}/\delta\approx\SI{27.6}{\per\second}\;.\]
    For $k^+=\SI{11.6}{\micro\Molar\tothe{-1}\second\tothe{-1}}$, this is the
    case when $c_0>\SI{2.4}{\micro\Molar}$. Due to possibly large exchange rates
    between sequestered and unsequestered G-actin~\cite{DeLaCruz2000} the bulk
    G-actin concentration might be higher than assumed (and therefore the
    effective polymerization rate lower than measured). However, our arguments
    above show that our results are valid even then.

\item {\bfseries G-actin polymerization rate:} We ran additional simulations
  with a polymerization rate of $k^+=\SI{5}{\per\micro\Molar\per\second}$ and
  $k^+=\SI{20}{\per\micro\Molar\per\second}$. In the latter case, the length of
  the filopodium is enhanced, but the reported effects due to steric
  interactions are still observable.  The filopodia become very short
  ($<\SI{300}{\nano\meter}$) in the former case, and the probability of
  partially collapsed filaments recovering (instead of collapsing completely) to
  the filopodial tip becomes significantly higher, which is to be expected
  because small positive height fluctuations are then sufficient to reach the
  upper stable configuration. We also ran simulations with a doubled
    G-actin concentration $c_0=\SI{20}{\micro\Molar}$ and an approximately
    halved G-actin polymerization rate of
    $k^+=\SI{5}{\per\micro\Molar\per\second}$, yielding filopodia of a length of
    about $\SI{1.4}{\micro\meter}$ with hollow bundles.
\item {\bfseries Diffusion constant:} To check the influence of the diffusion
  constant, we ran simulations with
  $D=\SI{2.5e6}{\nano\meter\tothe{2}\per\second}$. This also led to a
  significantly reduced filament height (which is to be expected), similar to
  the reduction in the polymerization rate.
\end{itemize}
Therefore we conclude that our observed actin bundle remodeling effects are
indeed robust against relatively large changes in the G-actin parameter values.

In an earlier work, Mogilner and Rubinstein~\cite{ref:mogilner05} analyzed the
mechanical stability of a filopodial bundle. At the maximum stable length, $L$,
the force enacted on the bundle via the membrane $F$ is equal to the buckling
force of the bundle~\cite{Landau1995}
\begin{equation}
  \label{eq:bucklingforce}
  F=\frac{\pi^2B_s}{4L^2}\;,
\end{equation}
where $B_s=E I$ is the bending stiffness, $E$ is Young's modulus and $I$ the
second moment of area (interactions with filament cross-linking molecules are
not considered). The resulting buckling length is $L=\pi\sqrt{B_s/4F}$. When
considering single actin filaments in the bundle as filled rods, the bending
force and hence the buckling length of different lateral arrangements of
filaments differ only in their second moment of area. A single filament with
radius $r_F$, cross-sectional area $A$ and distance $\vec{a}$ from the bundle
center contributes to the bundle's second moment of area via the Huygens-Steiner
theorem
\begin{equation*}
  I_s(a)=\int_A|\vec{r}-\vec{a}|^2 dA=\frac{\pi}{4}r_F^4+\pi r_F^2a^2\;.
\end{equation*}
Hence, a full bundle in an hexagonal arrangement with interfilament spacing $d$
with $19$ filaments has a second moment of area of
\begin{equation*}
  \begin{split}
    I_F&=I_s(0)+6I_s(d)+6I_s(2d)+6I_s\bigl(\sqrt{3}d\bigr) \\
    &=\pi r_F^2\biggl(\frac{19}{4}r_F^2+48d^2\biggr)\;.
  \end{split}
\end{equation*}
A hollow bundle in an hexagonal arrangement with all seven inner filaments
collapsed exhibits a second moment of area of
\begin{equation*}
  I_O=6I_s(2d)+6I_s\bigl(\sqrt{3}d\bigr)=\pi r_F^2\bigl(3r_F^2+42d^2\bigr)\;.
\end{equation*}
Hence, the buckling length ratio between hollow and intact
filopodial bundles is (assuming $d=\SI{13}{\nano\meter}$)
\[f=\sqrt{\frac{3r_F^2+42d^2}{\frac{19}{4}r_F^2+48d^2}}\approx0.93\;.\]
Therefore, the maximum stable length of a totally hollowed-out bundle is reduced
to $93\%$ of the intact bundle, i.e. the effect on the mechanical stability is
rather weak. If, on the other hand, the bundle is laterally coalesced into a
hexagonal arrangement by, for example, strong cross-linker interactions or
molecular motors, the second moment of area becomes
\begin{equation*}
  I_C=I_s(0)+6I_s(d)+5I_s\bigl(\sqrt{3}d\bigr)=\pi r_F^2\bigl(3r_F^2+21d^2\bigr)\;,
\end{equation*}
and the above-mentioned length ratio becomes $f\approx 0.66$, producing a much
larger effect because of the reduced cross-sectional area of the filament
bundle.

\section{MEAN-FIELD MODEL}
\label{sec:mean-field-model}
\begin{figure}[tbp]
  \centering
  \begin{tikzpicture}
    \begin{scope}[xscale=2.45, yscale=0.5]
      \draw[->] (0,4.3) -- node[above] {Filopodial height} +(3,0);
      \draw (0,3) -- (3,3) -- (3,0) -- (1,0) -- (1,1) -- (0,1);
      \draw (0,2) -- (2.5,2);

      \draw[blue,line width=0.1cm] (0,4) -- +(2.5,0);
      \draw[green!50!black,line width=0.1cm] (0,3.6) -- +(1,0);
      \draw[blue,line width=0.1cm] (0,3.3) -- +(2.5,0);

      \draw (1.25,2.5) node {\color{blue}$c_1$, $A_1$};
      \draw (0.5,1.5) node {\color{green!50!black}$c_2$, $A_2$};
      \draw (1.75,1) node {\color{red}$c_3$, $A_3$};
      \draw (2.75,2) node {\color{cyan}$c_4$, $A_4$};
      \draw[dashed] (0,1) -- (0,3);
      \draw[dashed] (1,1) -- node[right] {$k_1$} (1,2);
      \draw[dashed] (2.5,0) -- node[pos=0.3333,right] {$k_2$} (2.5,3);
      \draw (0,1) node [below] {$0$};
      \draw (1,0) node [below] {$L_1$};
      \draw (2.5,0) node [below] {$L_2$};
      \draw (3,0) node [below] {$L$};
      \draw (0,2) node [left] {$c_0$};
      \draw (current bounding box.north west) node [below right] {(a)};
    \end{scope}
    \draw (current bounding box.north west) +(3.25in, 0) rectangle (current bounding box.south west);
  \end{tikzpicture}

  \includegraphics[width=0.99\columnwidth,page=5]{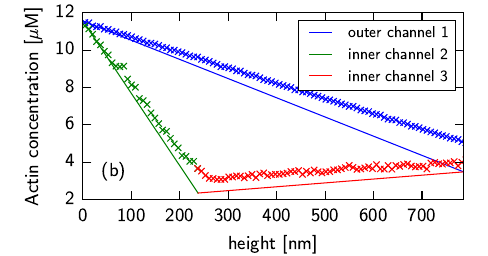}

  \includegraphics[width=0.99\columnwidth,page=6]{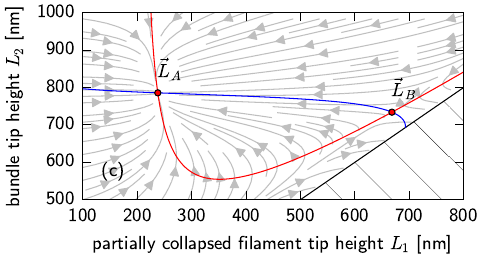}

  \caption{\label{fig:one-d-channels} (a) 
  {\it Sketch of the one-dimensional
    diffusion channels considered as part of the mean-field model for a
    filopodium with two distinct filament heights. The two channels exhibit
    {\normalsize $x$}-dependent piece-wise linear concentrations 
    {\normalsize $c_1$},
    {\normalsize $c_2$}, {\normalsize $c_3$} 
    and {\normalsize $c_4$}. The lower channel changes
    its cross-section at position {\normalsize $L_1$} due to the 
    presence of one or
    more filament tips. At position {\normalsize $L_2$}, 
    both channels merge and the
    cross-sectional area changes again with {\normalsize $A_1+A_3 < A_4$}. 
    The
    membrane is at {\normalsize $L=L_2+\SI{25}{nm}$}. 
    Polymerization of G-actin is
    implemented via sinks with strength {\normalsize $k_1$} and 
    {\normalsize $k_2$} at
    positions {\normalsize $L_1$} and {\normalsize $L_2$}, respectively.} 
    (b) {\it Mean-field
    G-actin concentration profiles in the three channels for meta-stable
    configuration} (C) 
    {\it displayed in Fig.\ $\ref{fig:example-config}$ together with
    data from stochastic simulations.} 
    (c) {\it Plot of the roots of the two stability
    conditions~$(\ref{eq:mf-stabilitycond})$ as a function 
    of {\normalsize $L_1$} and
    {\normalsize $L_2$} for the same configuration. 
    The points of filament stability
    where both conditions are true simultaneously are indicated by red dots. The
    light gray lines show the flow of the stability
    conditions~$(\ref{eq:mf-stabilitycond})$. The hatched area indicates the
    unphysical regime {\normalsize $L_1\ge L_2$}.}}
\end{figure}
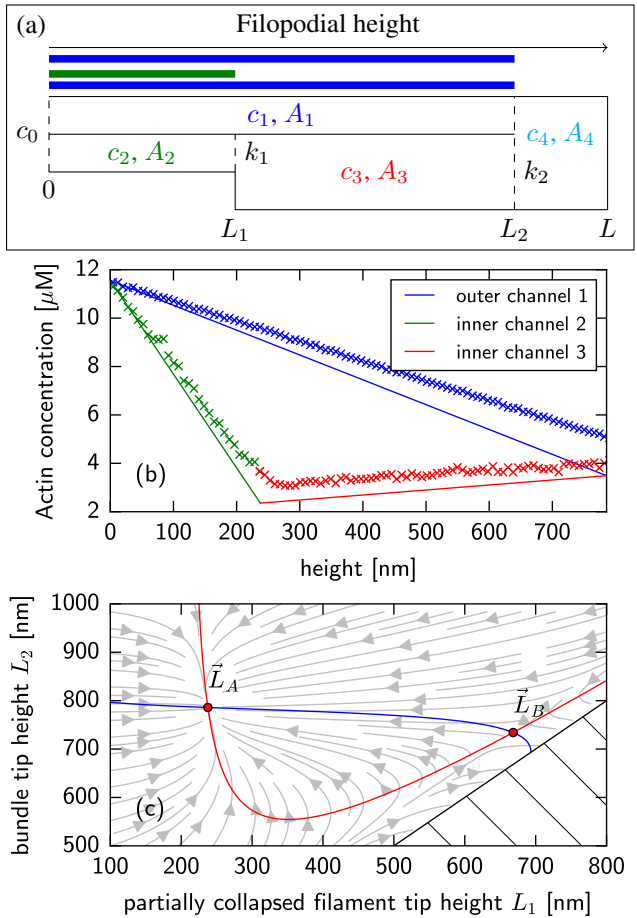

In order to provide deeper understanding into the processes that lead to
filament collapse inside the bundle's core, we next introduce the following
simplified model. We assume that G-actin diffusion occurs fast, such that linear
concentration profiles are quickly established along the filopodia axis. The
arrangement of filaments is such that G-actin cannot pass between two
neighboring filaments. Fig.\ \ref{fig:one-d-channels}(a) shows a sketch of this
channel configuration. Hence, there exist at least two separate channels for
diffusion in the filopodium: G-actin can diffuse along and {\it outside} of the
filament bundle, as well as via any inner space that opens up due to filament
collapse. This inner channel can then host filaments that are partially
collapsed, which yields a changing cross-sectional area of the diffusion
channel. The outer channel exhibits a cross-sectional area $A_1$. The area of
the inner lower channel is given by $A_2$, which widens to $A_3$ at the
partially-collapsed filaments tip position $L_1$. The two channels merge to a
single diffusion channel with area $A_4=\pi r_F^2$ at the bundle's tip $L_2$. In
a meta-stable state, the filament tips in the partially-collapsed level as well
as the bundle's top both consume just enough G-actin in order to hold their
height stable. The G-actin concentrations in the corresponding channels $c_1$,
$c_2$, $c_3$ and $c_4$ then follow the set of diffusion equations
\begin{gather}
  \label{eq:mf-diffeq}
  \begin{aligned}
    \frac{d^2 c_1}{dx^2}&=0,\;x\in[0,L_2] & \frac{d^2 c_2}{dx^2}&=0,\;x\in[0,L_1] \\
    \frac{d^2 c_3}{dx^2}&=0,\;x\in[L_1,L_2] & \frac{d^2
      c_4}{dx^2}&=0,\;x\in[L_2,L]\;,
  \end{aligned}
\end{gather}
with the conditions at the boundaries of the different channels
\begin{equation}
  \begin{split}
    c_1(0)&=c_2(0)=c_0\;, \\
    c_2(L_1)&=c_3(L_1)\;, \\
    c_1(L_2)&=c_3(L_2)=c_4(L_2)\;,
  \end{split}
  \label{eq:mf-diffcond}
\end{equation}
and the flux conditions across these boundaries
\begin{equation}
  \begin{split}
    &A_2\frac{d c_2}{dx}(L_1)-A_3\frac{d c_3}{dx}(L_1)=-\frac{k_1}{D}c_3(L_1)\;, \\
    &A_1\frac{d c_1}{dx}(L_2)+A_3\frac{d c_3}{dx}(L_2)=-\frac{k_2}{D}c_4(L_2)\;, \\
    &\frac{d c_4}{dx}(L)=0\;.
  \end{split}
  \label{eq:mf-diffcondflux}
\end{equation}
The coefficient of the reactive boundary condition describing the removal of
G-actin by the $N_1$ partially-collapsed filament tips is given by
$k_1=N_1k^+/N_A$, where $N_A$ is Avogadro's constant. Similarly, the removal of
molecules via the $N_2$ filaments at the bundle tip is implemented by a reactive
boundary condition at $L_2$ with the coefficient $k_2=f_m(N_2) N_2 k^+ /
N_A$.
The factor $f_m(N_2)$ stems from the steric interactions of G-actin with the
membrane at the top of the filopodium, and is calculated via
\begin{equation*}
  \label{eq:avg-fM}
  f_m(N_2)=\num{0.550}+\num{2.524e-2}N_2-\num{5.131e-4}N_2^2\;,
\end{equation*}
where the coefficients result from a fit of a polynomial of order 2 to data
generated in simulations and listed in Table~\ref{tab:membrane-factor}.
Eqs~(\ref{eq:mf-diffeq})-(\ref{eq:mf-diffcondflux}) can be written as a system
of linear equations for the concentrations at the filament tips $c_3(L_1)$ and
$c_1(L_2)$.
\begin{table}
  \caption{Membrane factor values obtained from simulations.}
  \label{tab:membrane-factor}
  \begin{tabular}{ll|ll}
    $N_2$ & $f_M(N_2)$ & $N_2$ & $f_M(N_2)$ \\
    \hline
    $12$ & $0.779$ & $16$ & $0.823$ \\
    $13$ & $0.792$ & $17$ & $0.831$ \\
    $14$ & $0.804$ & $18$ & $0.838$ \\
    $15$ & $0.814$ & $19$ & $0.845$ \\
    \hline
  \end{tabular}
\end{table}

Fig.\ \ref{fig:one-d-channels}(b) shows the resulting linear concentration
profiles in the respective channels for the example configuration (C) shown in
Fig.\ \ref{fig:example-config}, together with data obtained from our
simulations. As discussed above, filament heights are \hbox{(meta-)stable} when the
polymerization of G-actin with rate $k^+$ is counterbalanced by the effective
reduction in length due to the depolymerization reaction with rate $k^-$ and
retrograde flow with speed $v$. Thus, the stability criteria are given by
\begin{equation}
  f_m\,k^+c_1(L_2)\frac{A_1}{A_4}=k^-+\frac{v}{\delta}\;,\quad
  k^+c_3(L_1)=k^-+\frac{v}{\delta}\;,
  \label{eq:mf-stabilitycond}
\end{equation}%
where $c_1(L_2)$ and $c_3(L_1)$ are the concentrations of G-actin at the
filament tip positions $L_2$ and $L_1$ in channels 1 and 3, respectively. The
area ratio is necessary due to the geometry of the bundle. The effective
reduction of the polymerization rate at the $L_2$ tip position due to membrane
interaction is taken into account by the phenomenological factor $f_m$. These
criteria together with the solution from
Eqs.~(\ref{eq:mf-diffcond})-(\ref{eq:mf-stabilitycond}) describe two non-linear
curves in the $0<L_1<L_2$ sector of the plane spanned by $L_1$ and $L_2$. If
there exists a stable configuration, the two curves intersect in two points,
which are given by $\vec{L}_A=(L_{1,A},L_{2,A})$ and
$\vec{L}_B=(L_{1,B},L_{2,B})$. The point $\vec{L}_A$ is a stable configuration
within the limits of this model, while the point $\vec{L}_B$ is a
  saddle point that mediates the collapse of filaments from the bundle
tip. Fig.\ \ref{fig:one-d-channels}(c) shows the two curves and their
intersection points for configuration (C) shown in Fig.\
\ref{fig:example-config}. The light gray arrows indicate the vector
  field
\[\Biggl(\frac{f_m k^+ c_1(L_2)}{k^-+v/\delta}\frac{A_1}{A_4}-1,\frac{k^+ c_3(L_1)}{k^-+v/\delta}-1\Biggr)\;,\]
which corresponds to the flow of the filament stability
conditions~(\ref{eq:mf-stabilitycond}). In the stochastic simulation model, the
point $\vec{L}_B$ becomes quasi-stable. As soon as a fluctuation in the height
of a single filament causes it to fall approximately $\SI{100}{\nano\meter}$
below $L_{1,B}$, the single filament will collapse to the height $L_{1,A}$ and
the bundle tip will move to a length $L_{2,A}$.

The black dashed lines in Fig.\ \ref{fig:example-config} show the predicted
filament heights for three different meta-stable configurations, with very good
agreement with our simulation data. The observed overestimation of the height of
the partially collapsed filaments and the underestimation of the height of the
bundle tip stems from the inability of the theory to capture the more
complicated spatial structure inside a bundle. This is also the reason for the
discrepancy between the mean field concentrations and the data from stochastic
simulations in Fig.\ \ref{fig:one-d-channels}(b).

In order to find the parameter regimes in which meta-stable bundle
  configurations exist (rather than inner filaments simply collapsing
  completely), we extracted the mean-field stability boundary of the bundle
  configuration (C) in Fig. 2 in the manuscript via finding the value of
  $k^+(c_0)$ at which Eqs.~(\ref{eq:mf-stabilitycond}) start to have
  solutions. It is well described by the curve given as
  $c_0 k^+\approx \SI{63.8}{\per\second}$.  However, this only applies to this
  particular bundle configuration. When we repeat the same analysis 
  over all bundle
  configurations we observe in simulations with
  $D=\SI{5e6}{\nano\meter\tothe{2}\per\second}$ and
  $d=\SI{13}{\nano\meter}$ but varying $c_0$ and $k^+$, the minimum value is
  $\alpha=c_0 k^+\approx \SI{46.9}{\per\second}$. This value of $\alpha$ can be
  viewed as the upper limit for the phase boundary between immediate collapse of
  inner filaments ($c_0 k^+<\alpha$) and the existence of meta-stable filament
  heights ($c_0 k^+>\alpha$).

\section{DISCUSSION}

Our results show that the sterically hindered movement of free G-actin molecules leads to intriguing effects: When G-actin cannot pass between filaments in the filopodial shaft bundle, the polymerization of barbed ends of F-actin inside the bundle may not be sufficient to counteract retrograde flow. However, instead of the complete collapse of all filaments, novel meta-stable intermediate-height states emerge. Interior filaments that arise from an initial transient growth regime, initially have similar heights compared with the stable exterior filaments. However, height fluctuations will eventually drive single or multiple interior filaments below a critical stable height $L_{1,B}$, after which they collapse to a new meta-stable height $L_{1,A}$. We verified that this effect is present for different values of the interfilament spacing $d$ observed in experimental measurements of actin bundles. Our mean-field model enables us to approximately calculate the expected critical heights for a given configuration. As these new states are only meta-stable, the partially collapsed filaments will eventually collapse fully and disappear. This, in turn, opens up new diffusion channels for G-actin molecules, raising their concentration in the bundle interior and thereby enhancing the stability of any remaining interior filaments. Due to the raised supply of G-actin, the average height of the topmost filament tips is raised as well.

The configuration of the filament bundle at any given time depends on the  history of all prior filament collapses starting from the initial transient. Therefore, the ensemble of observed bundle configurations is highly diverse, with a small selection of configurations shown in Fig.\ \ref{fig:metastable-states}. Over time, the interior of the bundle becomes hollowed-out with some interior filaments collapsing partially or completely.  Thus, volume exclusion and the resulting change in the free diffusion of G-actin facilitates sculpting of the actin bundle inside filopodia. This reaction-diffusion sculpting mechanism adds complexity to the formin/capping protein mediated filament dynamics investigated in
reference~\cite{ref:zhuravlev09}.

Furthermore, we should consider the possibility that internal forces due to
filament cross-linking proteins or molecular motors may laterally constrict
bundles with collapsed internal filaments. Hence, as the filopodium ages, two
outcomes are likely: 1) After the bundle becomes hollowed-out due to
reaction-diffusion sculpting, it continues to remain hollow (for example,
because of weak cross-linking activity), experiencing only a small reduction of
its initial mechanical stability, as discussed above; 2) The bundle shrinks
radially due to inward internal forces collapsing the hollow cavity, which, in
turn, would lead to a strong reduction of bundle's mechanical stability. This
second scenario would imply an overall conical shape of aged filopodial bundles,
with a thicker portion position near the filopodial base and a thinner section
positioned at the filopodial tips, hence, explaining the corresponding common
observations of conically shaped actin filament bundles in superresolution and
fluorescence microscopy imaging of filopodia; see e.g. figure 1A in
reference~\cite{Suraneni2015} or figure 1C in reference
\cite{Svitkina2003}. Tapering of long filopodial protrusions has also been
reported in sea urchins~\cite{Miller1995}, in epithelial
cells~\cite{Khurana2011} and in plant cells~\cite{Grolig2014}.

\section*{AUTHOR CONTRIBUTIONS}
U.D. wrote and performed the simulations; U.D. and R.E. evaluated the computed
results; U.D., G.A.P. and R.E. designed the project and co-wrote the paper.

\section*{ACKNOWLEDGMENTS}
The research leading to these results has received funding from the European
Research Council under the European Community's Seventh Framework Programme
(FP7/2007-2013) / ERC grant agreement n$^o$ 239870. R.E. would like to thank the
Royal Society for a University Research Fellowship and the Leverhulme Trust for
a Philip Leverhulme Prize. This prize money was used to support research visits
of G.A.P. in Oxford. His work was also supported by NSF grant
CHE-1363081. R.E. and U.D. would like to thank the Isaac Newton Institute for
Mathematical Sciences, Cambridge, for support and hospitality during the
programme ``Stochastic Dynamical Systems in Biology: Numerical Methods and
Applications'' where work on this paper was undertaken. This work was supported
by EPSRC grant no EP/K032208/1. This work was partially supported by a grant
from the Simons Foundation.  U.D. was supported by a Junior Interdisciplinary
Fellowship via Wellcome Trust grant number 105602/Z/14/Z.

\bibliography{2015BIOPHYSJ.bib}

\end{document}